\documentclass[12pt,preprint]{aastex}
\usepackage{bbm}
\usepackage{natbib}
\usepackage{rotating}
\usepackage{amsmath}

\newcommand{\beq}{\begin{eqnarray}}
\newcommand{\eeq}{\end{eqnarray}}
\shorttitle{Dark Energy from SNe Ia and Strong Gravitational Lenses}
\shortauthors{Yue-Liang Wu, et al.}

\begin{document}

\title{ Dark Energy and Hubble Constant \\ From the Latest SNe Ia,
BAO and SGL}
\author{Qing-Jun Zhang and Yue-Liang Wu}

\affil{ Kavli Institute for Theoretical Physics China,
\\ Key Laboratory of Frontiers in  Theoretical Physics, \\ Institute of
Theoretical Physics, Chinese Academy of Science, \\ Beijing 100190,
China}


 \email{ylwu@itp.ac.cn}

\begin{abstract}
Based on the latest MLCS17 SNe Ia data provided by Hicken et al.
(2009), together with the Baryon Acoustic Oscillation (BAO) and
strong gravitational lenses (SGL), we investigate the dark energy
equation-of-state parameter for both constant $w$ and time-varying
$w(z)=w_0+w_az/(1+z)$ in the flat universe, and its correlation with
the matter density $\Omega_M$ and Hubble constant $h$. The
constraints from SNe data alone arrive at: (a) the best-fit results
are $(\Omega_M, w, h)=(0.358, -1.09, 0.647)$, while both $\Omega_M$
and $w$ are very sensitive to the difference $\Delta h =h-\tilde{h}$
of the Hubble constant deviating to the prior input
$\tilde{h}=0.65$; (b) the likelihoods of parameters are found to be:
$w = -0.88^{+0.31}_{-0.39}$ and $\Omega_M=0.36^{+0.09}_{-0.15}$,
which is consistent with the $\Lambda \rm CDM$ at $95\%$ C.L.; (c)
the two parameters in the time-varying case are found to be $(w_0,
w_a)=(-0.73^{+0.23}_{-0.97}, 0.84^{+1.66}_{-10.34})$ after
marginalizing other parameters; (d) there is a clear degeneracy
between constant $w$ and $\Omega_M$, which depresses the power of
SNe Ia to constrain both of them; (e) the likelihood of parameter
$w_a$ has a high non-Gaussian distribution; (f) an extra restriction
on $\Omega_M$ is necessary to improve the constraint of the SNe Ia
data on ($w_0$, $w_a$). A joint analysis of SNe Ia data and BAO is
made to break the degeneracy between $w$ and $\Omega_M$, and it
provides a stringent constrain with the likelihoods: $w =
-0.88^{+0.07}_{-0.09}$ and $\Omega_M=0.29^{+0.02}_{-0.03}$.  For the
time-varying $w(z)$, it leads to the interesting maximum likelihoods
$w_0 = -0.94$ and $w_a = 0$. When marginalizing the parameters
$\Omega_M$ and $h$, the fitting results are found to be $(w_0,
w_a)=(-0.95^{+0.45}_{-0.18}, 0.41^{+0.79}_{-0.96})$. After adding
the splitting angle statistic of SGL data, a consistent constraint
is obtained $(\Omega_M, w)=(0.298, -0.907)$ and $(w_0, w_a)$ is
further improved to be $(w_0, w_a) = (-0.92^{+0.14}_{-0.10},
0.35^{+0.47}_{-0.54})$, which indicates that the phantom type models
are disfavored.
\end{abstract}



\section{INTRODUCTION}

  Analysis of the distance modulus versus redshift relation of type
Ia supernova (SNe Ia) provides a direct evidence that the universe
expansion is accelerating in the last few billion years(e.g., Riess
et al. 1998, 2004, 2007; Wood-Vasey et al. 2007; Kowalski et al.
2008; Hicken et al. 2009; Biswas and Wandelt1 2009). This cosmic
image is also supported by many other cosmological observations,
like the Cosmic Microwave Background(CMB)(Hinshaw et al. 2009;
Komatsu et al. 2009), the Baryon Acoustic Oscillation (BAO)
measurement (York et al. 2000; Eisenstein et al. 2005) and the weak
gravitational lenses (Weinberg and Kamionkowski 2002;
 Zhan and Knox 2006). Based on the Friedmann equation, the
acceleration can be explained through introducing a negative
pressure component in the universe, named dark energy, which is
nearly spatially uniform distribution and contributes about $2/3$
critical density of universe today. To reveal the property of dark
energy, the most of studies, either theoretical models or experiment
data analysis, are focused on its equation-of-state parameter
$w=p/\rho$(Albrecht et al. 2006). Here we shall utilize the latest
SNe Ia data provided by Hicken et al. (2009) with using MLCS17 light
curve fitter, together with the splitting angle statistic of strong
gravitational lenses(SGL; Zhang et al. 2009) and the baryonic
acoustic oscillations (BAO; Eisenstein et al. 2005) to investigate
the constraint for the parameter $w$ of dark energy in the flat
cosmology.  The influences of Hubble constant h and matter density
$\Omega_M$ on the fitting results  are carefully demonstrated.


By far, all observed data are consistent with the $\Lambda {\rm
CDM}$ cosmology, with dark energy in the form of a cosmological
constant $\Lambda$. However, this model raises theoretical problems
related to the fine tuned value (see e.g. Padmanabhan 2003). Many
other theoretical models, like quintessence models and phantom
model(Ratra and Peebles 1988; Caldwell, Dav{\'e}, and Steinhardt
1998; Caldwell 2002) , reveal that dark energy might be a dynamical
component and evolves with time. It is usual to parametrize dark
energy as an ideal liquid  with its equation-of-state(EOS) parameter
$w(z)=w_0 + w_a \; z/(1+z)$ (Chevallier et al. 2001; Linder 2003).
Conveniently, it includes the case of a constant EOS with ($w_0 =
w,w_a = 0$), and the $\Lambda {\rm CDM}$ model ($w_0 = -1,w_a = 0$).
Then theoretical models can be classified in a phase diagram on the
($w_0,w_a$) plane (see e.g. Barger et al. 2006; Biswas and Wandelt
2009). Thus the accurate measurement of the parameters ($w_0, w_a$)
is helpful for testing a certain theoretical models. The current
allowed regions of ($w_0, w_a$) given by the most observations or
their combinations remain surrounding the crucial point ($w_0 =-1,
w_a=0$), which is the common point in the phase diagram of different
classified models. Therefore, the final judgment of models can not
be made and more careful works are still needed.

In practice, the measurements of other cosmological parameters, like
Hubble constant $h$ and matter density parameter $\Omega_M$, shall
affect the determination of  the EOS parameter $w$ of dark energy.
Hubble constant $h$ gives the rate of recession of distant galaxies
and scales the expansion of the present universe. Parameter
$\Omega_M$ describes the present matter density relative to the
critical density, which is essential to understand the structure
formation. In the flat cosmology, $\Omega_M$ is directly related to
the present density of dark energy by $\Omega_{DE} = 1 - \Omega_M$.
The precise measurements on the two cosmological parameters are
still processing. The results from WMAP five-year observations are
$h = 0.719 \pm 0.03$ and $\Omega = 0.258 \pm 0.03$ in the $\Lambda
\rm CDM$ cosmology, respectively(Hinshaw et al. 2009). The SNe Ia
also have been used to determine the Hubble constant, calibrated
with their peak luminosities or Cepheid variables in nearby
galaxies(Schaefer 1996; Sandage, et al. 1996; Branch 1998; Gibson
2001; Sandage et al. 2006; Riess et al. 2009). Because the Hubble
constant  degenerates with the distance modulus of SNe Ia, the
correct value of the absolute magnitude of the fiducial SN Ia or
Hubble constant is not relevant for the dark energy analysis which
only make use of differences between SNe Ia magnitudes(Riess et al.
2004). In this case, the value $h = 0.65$ is often used as a prior
input(Hicken et al. 2009). However, we can still investigate the
sensitivity of the analysis results on the change of the Hubble
constant, which is not presented in the literatures.

The SNe Ia has homogeneity and extremely high intrinsic luminosity
of peak magnitude and thus is widely used to measure the
cosmological parameters(e.g. Riess et al. 2004; Wood-Vasey et al.
2007; Kowalski et al. 2008; Hicken et al. 2009). With given density
of dark energy in the universe today $\rho(z=0)$, the change of
equation-of-state parameter $w$ will bring the change of its density
$\rho(z)$ at the redshift $z$ and then the distance $d(z)$.
Inversely, measurement of the redshift $z$  of SNe Ia and
corresponding distance $d(z)$ can constrain the dark energy. In
spite of the high accuracy of SNe Ia measurement, its potential of
constraining dark energy is not very strong due to the degeneracy of
$w$ and $\Omega_M$. Therefore, a combining analysis with other
observations is often useful.

We will also use the  summary parameters of the baryonic acoustic
oscillations as reported in previous studies(Eisenstein et al.
2005). The large-scale correlation function of a large sample of
luminous red galaxies has been measured in the Sloan Digital Sky
Survey and a well-detected acoustic peak was found to provide a
standard ruler by which the absolute distance of $z=0.35$ can be
determined with $5\%$ accuracy,  independent on the Hubble constant
$h$. This ruler is a $\Omega_M$ prior and can be used to constrain
dark energy(e.g. Porciani and Madau 2000; Huterer and Ma 2004; Chae,
2007). We shall also use strong gravitational lensing statistic
observation, which provides us a useful probe of dark energy of the
universe. Mainly through the comoving number density of dark halos
described by Press-Schechter theory and the background cosmological
line element, dark energy affects the efficiency with which
dark-matter concentrations produce strong lensing signals. Then
through comparing the observed number of lenses with the theoretical
expected result as a function of image separation and cosmological
parameters, it enables us to determine the allowed range of the
parameter $w$. The constraint process also depends on the density
profile of dark halos. Here we shall use the two model combined
mechanism to reproduce the observed curve of lensing probability to
the image splitting angle (Sarbu, Rusin and Ma 2001; Li and Ostriker
2002; Zhang et al. 2009). The redshift of ${\rm CMB}$ is above
$1000$ and far larger than $1$, and there is no other observation to
fill up this redshift gap, thus we would not adopt the CMB data in
the present analysis and limit our study on dark energy to the
redshift region of $z \sim 1$, which is the characteristic redshift
scale of SNe Ia, BAO and SGL statistic.

In our recent work(Zhang et al. 2009), we have present the
constraint on the dark energy from the SGL splitting angle
statistic. In this paper, by taking the latest analyzed SNe Ia
data(Hicken et al. 2009), the baryonic acoustic
oscillations(Eisenstein et al. 2005) and the CLASS statistical
sample(Browne et al.2003), we shall make an joint analysis to
constrain the dark energy equation of state parameter $w$ and the
matter density $\Omega_M$, and to study the possible influence of
the Hubble constant $h$. Two model independent assumptions of dark
energy are considered: constant $w$ and the time-varying
parameterization $w(z) = w_0 + w_a \, z / (1 + z)$. We mainly
highlight two issues which have not previously been illuminated.
First, we carefully study, based on the latest SNe Ia data, the
constraints for the dark energy EOS $w(z)$ and the influences of the
matter density $\Omega_M$ and Hubble constant $h$. Second, we
investigate the joint analysis of SNe Ia data, BAO and SGL statistic
in detail. Our paper is organized as follows: Sect. 2 shows the
constraint by the latest SNe Ia data on dark energy. Sect. 3
describes the joint analysis of the SNe Ia, the BAO and the SGL
statistic, more stringent constraints on dark energy and matter
density are resulted, and the influence of Hubble constant is
explicitly demonstrated. The conclusions are presented in the last
section.

\section{DARK ENERGY CONSTRAINTS BY THE LATEST SNe Ia DATA}

As the standard candles of the cosmology, the SNe Ia is used to
study the geometry and dynamics of the universe with redshift $z\leq
1.7$. In determinations of cosmological parameters about the
accelerating expand and dark energy, the SNe Ia remains a key
ingredient. In 1998, the SNe Ia measurement provided the first
direct evidence for the presence of dark energy with the negative
pressure. Then many SN Ia observations have been done and the total
number of SNe Ia sample increases quickly. The SN Ia compilations
are often consist of high-redshift $(z \simeq 0.5)$ data set and
low-redshift $(z \simeq 0.05)$ sample at the same time (e.g. Riess
et al. 1998; Perlmutter et al. 1999; Wood-Vasey et al. 2007). When
combining several independent group's SNe Ia data sets into one
compilation, the consistent  analysis method of light curves and the
selection of supernova are crucial. For a certain sample, the
different light curve fitter and corresponding different selection
of supernova can lead to different constraints on the  cosmological
parameters(e.g. Hicken et al. 2009).

The fitting results of cosmological parameters from different SN Ia
compilations show an obvious difference which could be huge in some
cases. For example, the sample provided by Wood-Vasey et al. (2007)
led a cosmology with near zero matter component $\Omega_M \approx
0$, while the best-fit result of $\Omega_M$  by Riess et al. (2004)
was found to be $\Omega_M \sim 0.5$. The fitting constant $w$ of
dark energy for the former sample was found to be $w \sim -0.75$ and
for the latter sample, it was found to be $w \sim -2.0$. On the
other hand, it is obvious that larger sample of SNe Ia is more
powerful for constraining cosmological parameters. Therefore it is
useful and attractive to adopt the SNe Ia data as many as possible.
There is a conventional method to combine several group's SNe Ia
compilations, namely, by introducing an extra nuisance parameter in
the $\chi^2$ statistic of every used SNe Ia sample and marginalizing
them over in the fit, all $\chi^2$ statistics of samples can be
summed into one total statistics (see e.g. Barger et al. 2006). The
nuisance parameters are considered as analysis-dependent global
unknown constants in the distances and are degenerate with the
Hubble constant $h$. Although this combined mechanism is wildly
adopted, the so-called analysis-dependent unknown constant is just
an averaged effect of analysis-dependent uncertainties.

For a certain SNe Ia, its `measuring results' from independent
groups often have visible differences from disparate light curve
fitting functions and analysis procedures. Riess et al. (2004) and
Wood- Vasey et al. (2007) provided 157 golden sample and 162 SNe Ia
sample, respectively. The two samples both adopt MLCS2k2 light-curve
fitter and  have 39 common SNe Ia, every SNe Ia of which has two
sets of  different redshift $z$ and distance moduli $\mu$. From the
$39$ common data, we can compare the two data sample directly and
then obtain the distance uncertainty $\sigma_{o,\mu}=0.21$ and
redshift uncertainty $\sigma_{o,z} = 0.00029$. Therefore it is clear
that the analysis-dependent uncertainties from different group are
quite large. Partially to solve this problem, Kowalski et al. (2008)
provided the Union data set, a compilation of 307 SNe Ia discovered
in different surveys. The heterogeneous nature of the data set have
been reflected and all SNe Ia sample are analyzed with the same
analysis procedure.

In the Union data set, all SNe Ia light curve are fitted by using
the spectral-template-based fit method of Guy et al. (2005) (also
known as SALT). There are other light curve fitters used in
literatures, such as SALT2 (Guy et al. 2007), MLCS2k2 (Jha, Riess,
and Kirshner 2007) with $R_V = 3.1$ (MLCS31) and MLCS2k2 with $R_V =
1.7$ (MLCS17). Hicken et al. (2009) compared these light curve
fitters and found that SALT produces high-redshift Hubble residuals
with systematic trends versus color and larger scatter than MLCS2k2,
and MLCS31 overestimates host-galaxy extinction while MLCS17 does
not. For a certain SNe Ia, the analysis outcomes of different light
curve fitters are not equal. Here we choose the SNe Ia compilation
provided by using MLCS17 light curve fitter with the best cuts $A_V
\leq 0.5$ and $\Delta < 0.7$ to constrain the dark energy.

In the flat universe, the Friedmann equation are given by
 \beq
 H^2(z) &=& H_0^2 \left[ \Omega_M (1+z)^3 + \Omega_{DE}(z)\right]
 \nonumber \\
\Omega_{DE}(z) &=& \left\{\begin{array}{ll}(1 - \Omega_M) (1
+z)^{3(1+w)} & for \;\;  constant \; w \,,\\
            (1 - \Omega_M) (1
+z)^{3(1+w_0+w_a)}e^{-3w_az/(1+z)}
        \hspace{0.5cm} & for \;\; w(z)=w_0 + w_a \, {z\over1+z} \,,\\
   \end{array} \label{eq:hub}
   \right.
 \eeq with Hubble constant $H_0 = 100 \; h \;{\rm km \;s^{-1}\;
 Mpc^{-1}}$.  The influence of cosmological parameter $w$ is focused on the
 dark energy density $\Omega_{DE}(z)$ and then the Luminosity distance
$d_L$, which is defined as
 \beq
 d_L(z) = (1+z) \int^z_0 {dz' \over H(z')} \label{eq:lumdis}
 \eeq

Analysis of the distance modulus versus redshift relation of SNe Ia
can give us the information about the cosmological parameters.
Distance estimates of SNe Ia are derived from the luminosity
distance, $d_{L} = ({{\cal L} / 4 \pi {\cal F}})^{1/2}$ where ${\cal
L}$ and ${\cal F}$ are the intrinsic luminosity and observed flux of
the SNe Ia, respectively. It is usual to introduce the apparent
magnitude $m$  and absolute magnitude $M$. The apparent magnitude
$m$ of a celestial body is a measure of its brightness as seen by an
observer on Earth and it is defined as $m = {\rm
log}_{\sqrt[5]{100}} ({\cal F}/{\cal F}^0) = -2.5 \, {\rm
log}_{10}({\cal F}/{\cal F}^0)$, where ${\cal F}^0$ is a reference
flux, which is the zero point by definition(used to be the observed
flux of Vega star). The absolute magnitude $M$ of a celestial body
outside of the solar system is defined to be the apparent magnitude
it would have if it were $10$ parsecs away, namely
 \beq
  M &\equiv& m(10{\rm pc})\nonumber \\
    &=& -2.5 \,{\rm log}_{10}[{\cal F}(10{\rm pc})/{\cal F}^0]=-2.5 \, {\rm
        log}_{10}[\frac{{\cal F}}{ {\cal F}^0} \,{d_L^2 \over ({\rm Mpc})^2} 10^{10}] \nonumber \\
    &=& m-2.5 \, {\rm log} \left[{d_L^2 \over ({\rm Mpc})^2}\right] -25
 \eeq
From the definition of the distance moduli $\mu=m-M$, we have
 \beq
 \mu=5\log d_L/{\rm Mpc}+25. \label{eq:dismod}
 \eeq
Using Equations (\ref{eq:hub}), (\ref{eq:lumdis}) and
(\ref{eq:dismod}), we can relate the parameter $w$ with the measured
redshift $z$ and distance moduli $\mu(z)$ of SNe Ia data. Define
$\tilde{d_L} = H_0 d_L$, then we get
 \beq
 \mu + 5\log H_0= 5\log \tilde{d_L} + 25. \label{eq:dismod1}
 \eeq
The right side of the above formula doesn't depend on the Hubble
constant and $h$ directly degenerates with the distance modula
$\mu$. For the dark energy analysis, the SNe Ia data often use the
certain value of $\tilde{h}=0.65$ as the prior input and only make
use of the differences of the logarithmic SNe Ia distances. However,
we can still investigate the influence on the statistic analytic
results when changing the Hubble constant $h$. Here $\Delta h =
h-\tilde{h}=h-0.65$ is equivalent to the difference between the
input Hubble constant and the real Hubble constant in the universe.
Therefore although the certain value of the parameter $h$ loses its
physics meaning, $\Delta h$ is still a physics quantity. The
likelihood for the parameters ($\Omega_M, w, h$) can be determined
from a $\chi^2$ statistic
 \beq
 \chi^2(\Omega_M,w, h) = \sum_i { (\mu^T_i(z_i;\Omega_M, w, h)-\mu^O_i)^2\over \sigma_i^2}
 \label{eq:snechi2}
 \eeq
where subscript $i$ denotes the $i$th SNe Ia data and $\sigma_i$ is
the observed uncertainty. $\mu^O$ and $\mu^T$ are the observed and
theoretical distance moduli, respectively.

Let us first discuss the constraints for the constant $w$ case.
Using the Powell minimization method(Press, et al 1992), we minimize
the likelihood function of the three parameters $(\Omega_M, w, h)$
and find that the coordinate of the best-fit point is $(\Omega_M, w,
h) = (0.358, -1.09, 0.647)$. The fitted result of the Hubble
constant $h$ is very close to the prior input value $\tilde{h}=0.65$
of Hicken et al. (2009) . It is noted that the value of $\Omega_M$
is large, in comparison with $\Omega_M = 0.258$ of the concordance
cosmology provided by WMAP five year data(Komatsu et al. 2009).

\begin{figure}
\epsscale{0.8} \plotone{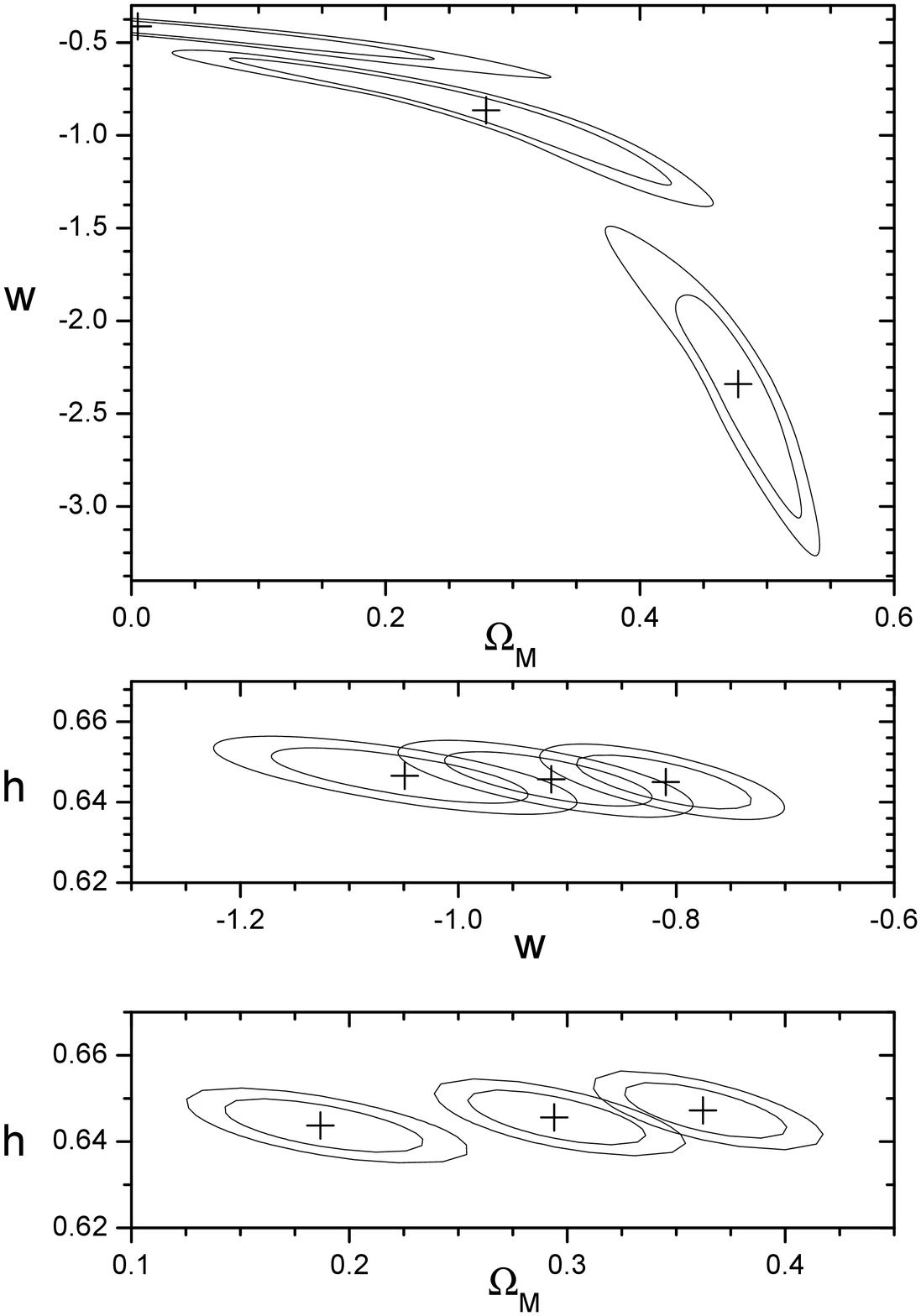}
\figcaption{\small The constraint results on the cosmological
parameters ($\Omega_M, w$), ($h, w$) and ($h, \Omega_M$) from top to
bottom,  respectively. In the upper panel, three pairs of contours
are for the given Hubble constants $h = 0.62$, $0.645$, and $0.67$
from top to bottom, respectively. The crosshairs mark the best fit
points $(\Omega_M, w)=(0.01, -0.41)$, $(0.28, -0.87)$ and $(0.48,
-2.34)$ from top to bottom. In the middle panel, three pairs of
contours are for the given $\Omega_M = 0.35$, $0.30$, and $0.25$
from left to right, respectively. The crosshairs mark the best fit
points $(w, h)=(-1.05, 0.647)$, $(-0.91, 0.646)$ and $(-0.81,
0.645)$ from left to right. In the lower panel, three pairs of
contours are for the given $w = -0.7$, $-0.9$, and $-1.1$ from left
to right, respectively. The crosshairs mark the best fit points
$(\Omega_M, h)=(0.19, 0.644)$, $(0.29, 0.646)$ and $(0.36, 0.647)$
from left to right. \label{fig:SNe1} }
\end{figure}

To see the correlations among the three parameters $(\Omega_M, w,
h)$ and their influences on the fitting results, we consider three
different combination fittings which are shown in figure
\ref{fig:SNe1}. From top to bottom in figure \ref{fig:SNe1}, we show
the fitting results of $(\Omega_M,w)$, $(w,h)$ and $(\Omega_M, h)$
with other parameter being varying in the resonable range. In every
panel, three pairs of contours are presented with a given
cosmological parameter: for the upper panel, the deviation of the
Hubble constant to the prior input $\Delta h = h - 0.65$ is taken to
be: $\Delta h = -0.03$, $\Delta h = -0.005$ and $\Delta h = 0.02$,
which are corresponding to $h = 0.620$, $0.645$, and $0.670$, from
top to bottom respectively; for the middle panel, the matter density
$\Omega_M$ is taken to be: $\Omega_M = 0.35$, $0.30$, and $0.25$
from left to right, respectively; for the lower panel, the constant
$w$ of dark energy is taken to be: $w = -0.7$, $-0.9$, and $-1.1$
from left to right, respectively. The crosshairs in every panel mark
the best fit points: for the upper panel, $(\Omega_M, w)=(0.01,
-0.41)$, $(0.28, -0.87)$ and $(0.48, -2.34)$ for $h = 0.620$,
$0.645$, and $0.670$; for the middle panel, $(w, h)=(-1.05, 0.647)$,
$(-0.91, 0.646)$ and $(-0.81, 0.645)$ for $\Omega_M = 0.35$, $0.30$,
and $0.25$; for the lower panel, $(\Omega_M, h)=(0.19, 0.644)$,
$(0.29, 0.646)$ and $(0.36, 0.647)$ for $w = -0.7$, $-0.9$, and
$-1.1$. It can be seen that the fitting results for $(\Omega_M,w)$
from the SNe Ia data is very sensitive to the value of Hubble
constant $h$: when $h$ increases, the fitting $\Omega_M$($w$)
increases(decreases) very rapidly.

\begin{figure}
\epsscale{0.8} \plotone{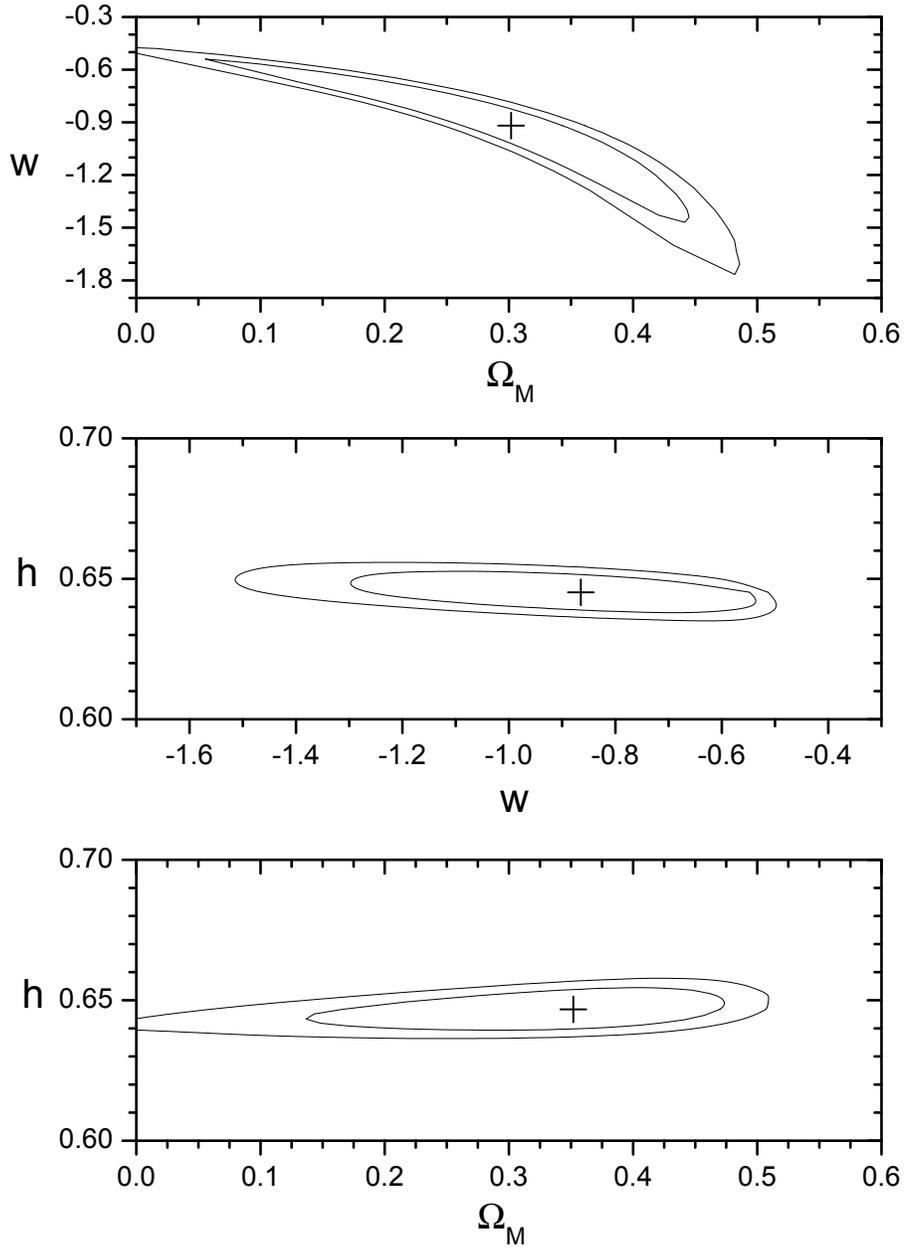}
\figcaption{\small 68\% C.L. and 95\% C.L.  allowed regions of
($\Omega_M, w$), ($h, w$) and ($h, \Omega_M$) respectively. The
crosshairs in three panels markThe best-fit points are $(\Omega_M,
w)=(0.30, -0.92)$, $(w, h)=(-0.86, 0.645)$ and $(\Omega_M, h)=(0.35,
0.647)$. from top to bottom. \label{fig:SNe2} }
\end{figure}

\begin{figure}
\epsscale{0.8} \plotone{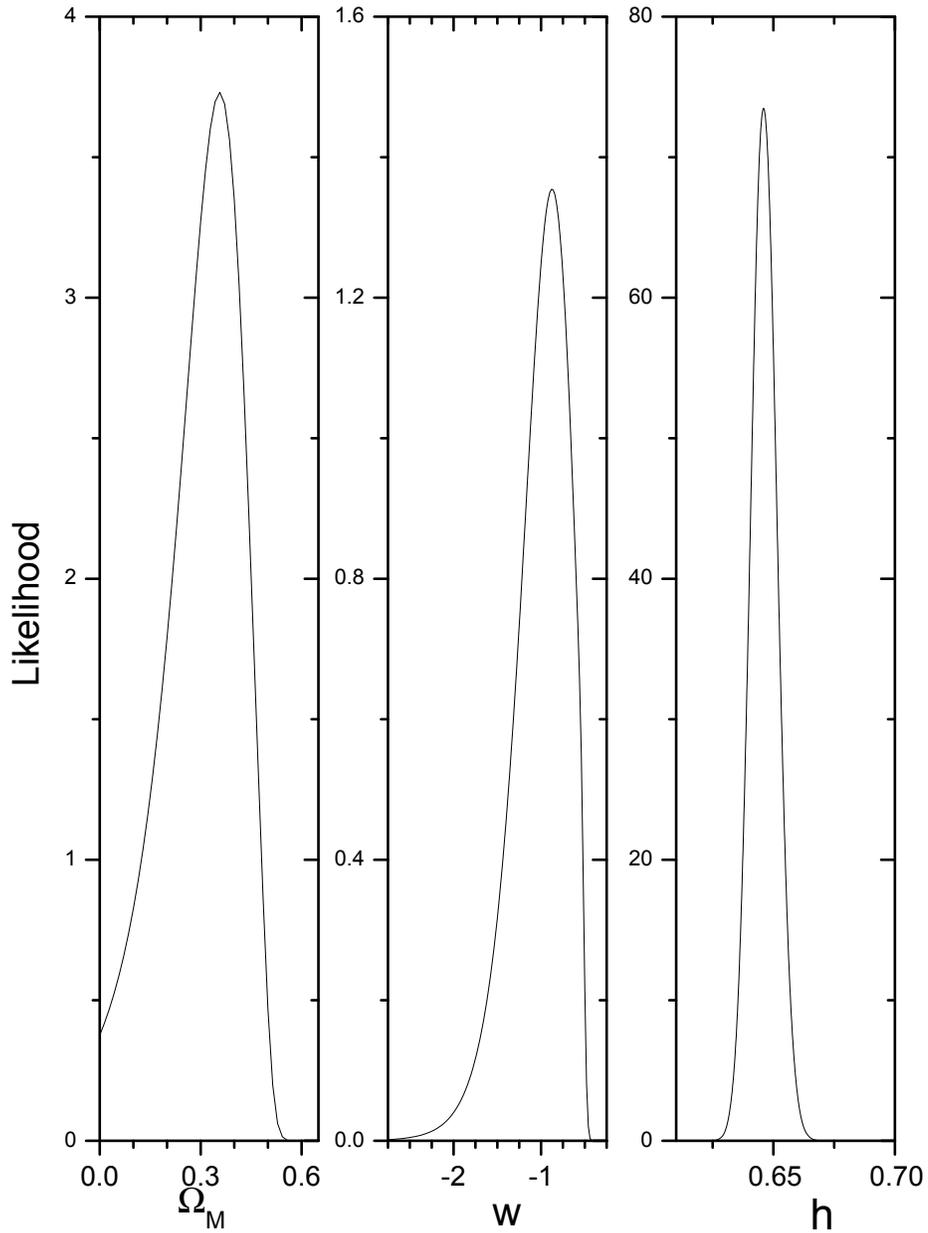}
\figcaption{\small The likelihoods of the parameters $\Omega_M$, $w$
and $h$. The maximum likelihood points are located at
$\Omega_M=0.36$, $w=-0.88$ and $h=0.65$, respectively.
\label{fig:SNe3}}
\end{figure}

To investigate the constraint on the interested parameters, it is
usual to marginalize other parameters. Figure \ref{fig:SNe2} shows
the $(\Omega_M,w)$, $(w,h)$ and $(\Omega_M, h)$ contours from top to
bottom, respectively. The unshown parameter in every panel has been
marginalized through integrating it by applying the likelihood
function $L = { \rm exp}(-\chi^2/2)$. The best-fit points are
$(\Omega_M, w)=(0.30, -0.92)$, $(w, h)=(-0.86, 0.645)$ and
$(\Omega_M, h)=(0.35, 0.647)$. The $(\Omega_M,w)$ contours have been
shown in figure 4 of paper by Hicken et al. 2009, and our
corresponding results are almost identical with theirs.  Here we
shall carefully discuss the information from the figures: (a) there
is an obvious degeneracy of constant $w$ and $\Omega_M$, which
depresses the power of SNe Ia to constrain both of them; (b) the
fitting results are consistent with the $\Lambda {\rm CDM}$
cosmology at $95\%$ C.L.. Comparing with the results obtained from
the Union data set (Kowalski et al. 2008), we notice that the
confidence regions in our present consideration are similar to those
of the Union data set. Figure \ref{fig:SNe3} shows the likelihoods
of parameter $\Omega_M$, $w$ and $h$, in which the maximum
likelihood points are located at $\Omega_M=0.36$, $w=-0.88$ and
$h=0.646$, respectively. It is interesting to notice that the
parameter $w$ is restricted to be from $-2.0$ to $-0.5$ and
$\Omega_M$ is less than $0.5$. Again, our fitted Hubble constant $h$
is consistent with the prior input value $\tilde{h}=0.65$ and its
allowed region is very narrow.

\begin{figure}
\epsscale{0.8} \plotone{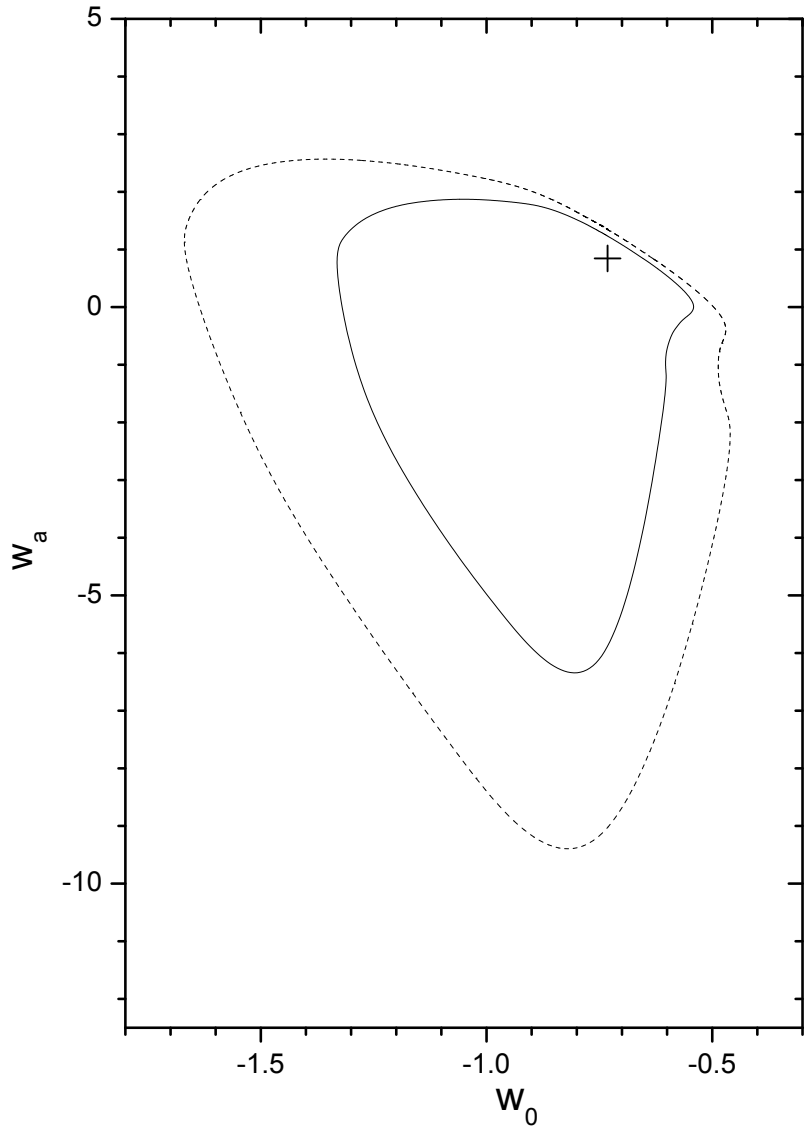}
\figcaption{\small The ($w_0, w_a$) contours of SNe Ia data alone.
The crosshairs mark the best-fit point ($w_0, w_a)=(-0.73,0.84)$.
\label{fig:SNe4} }
\end{figure}

\begin{figure}
\epsscale{0.8} \plotone{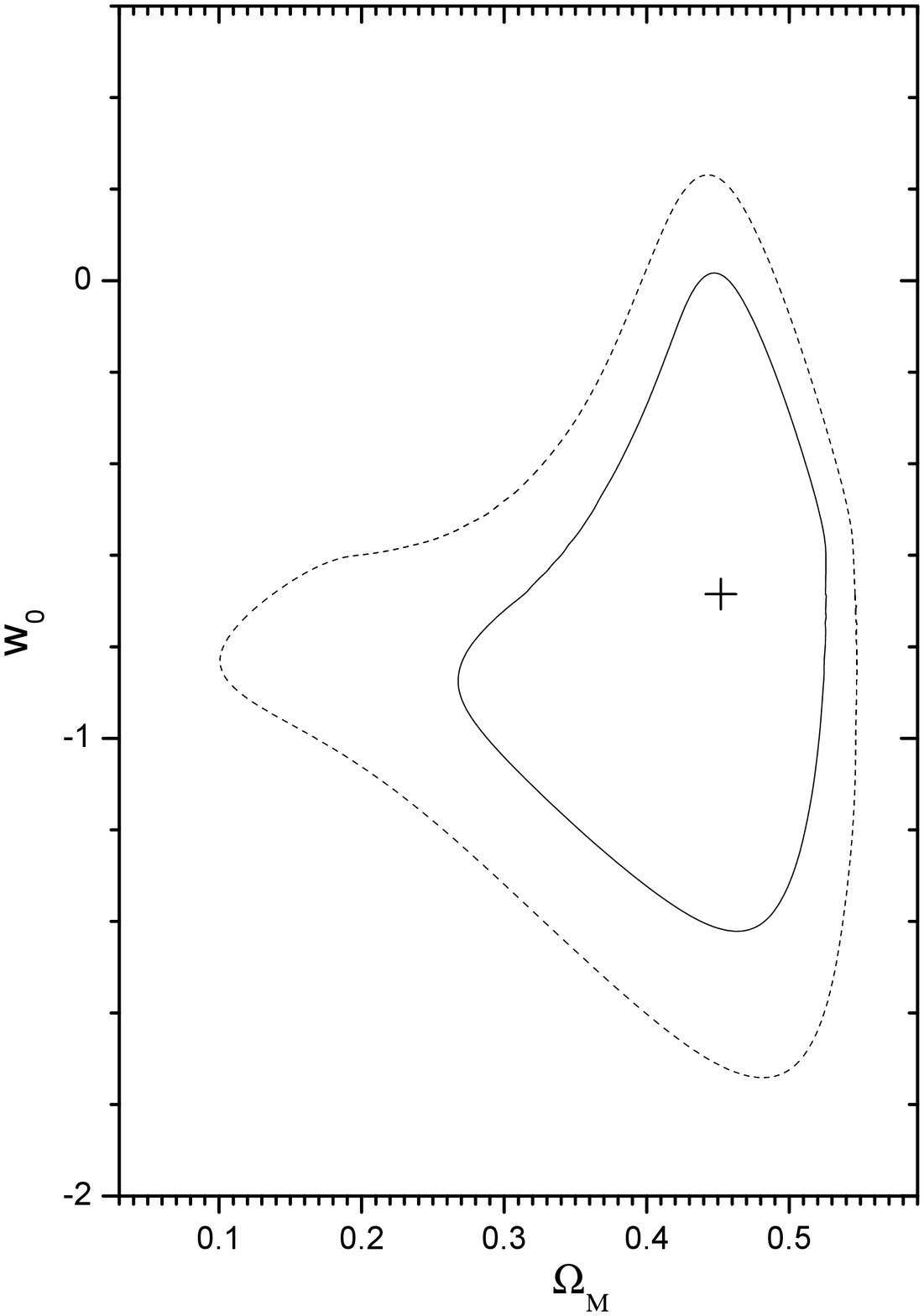}
\figcaption{\small The ($\Omega_M, w_0$) contours of SNe Ia data
alone. The crosshairs mark the best-fit point ($\Omega_M,
w_0)=(0.45, -0.68)$.  \label{fig:SNe5} }
\end{figure}

\begin{figure}
\epsscale{0.8} \plotone{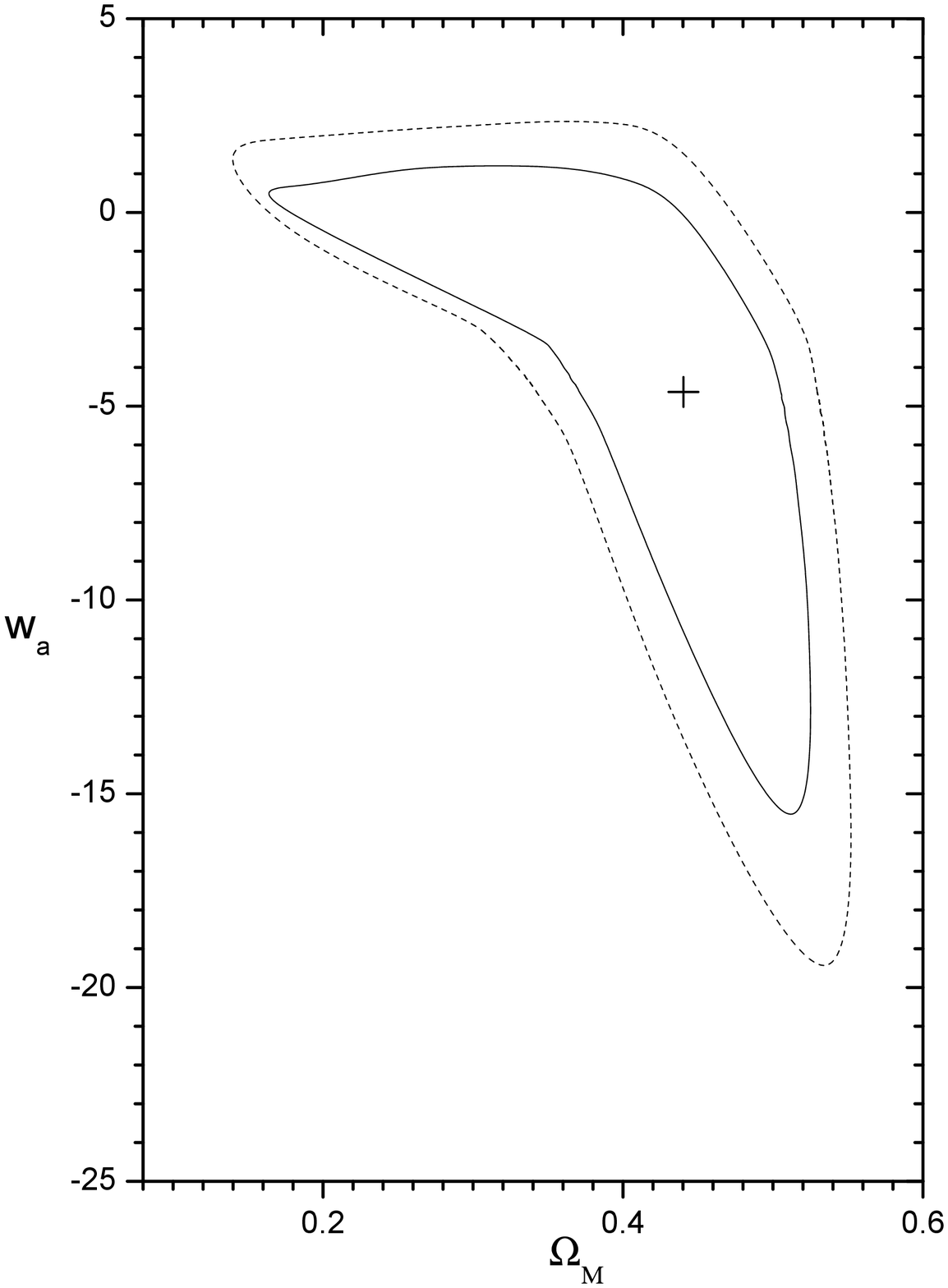}
\figcaption{\small The ($\Omega_M, w_a$) contours of SNe Ia data
alone. The crosshairs mark the best-fit point ($\Omega_M,
w_a)=(0.44, -4.63)$. \label{fig:SNe6}}
 \end{figure}

\begin{figure}
\epsscale{0.8} \plotone{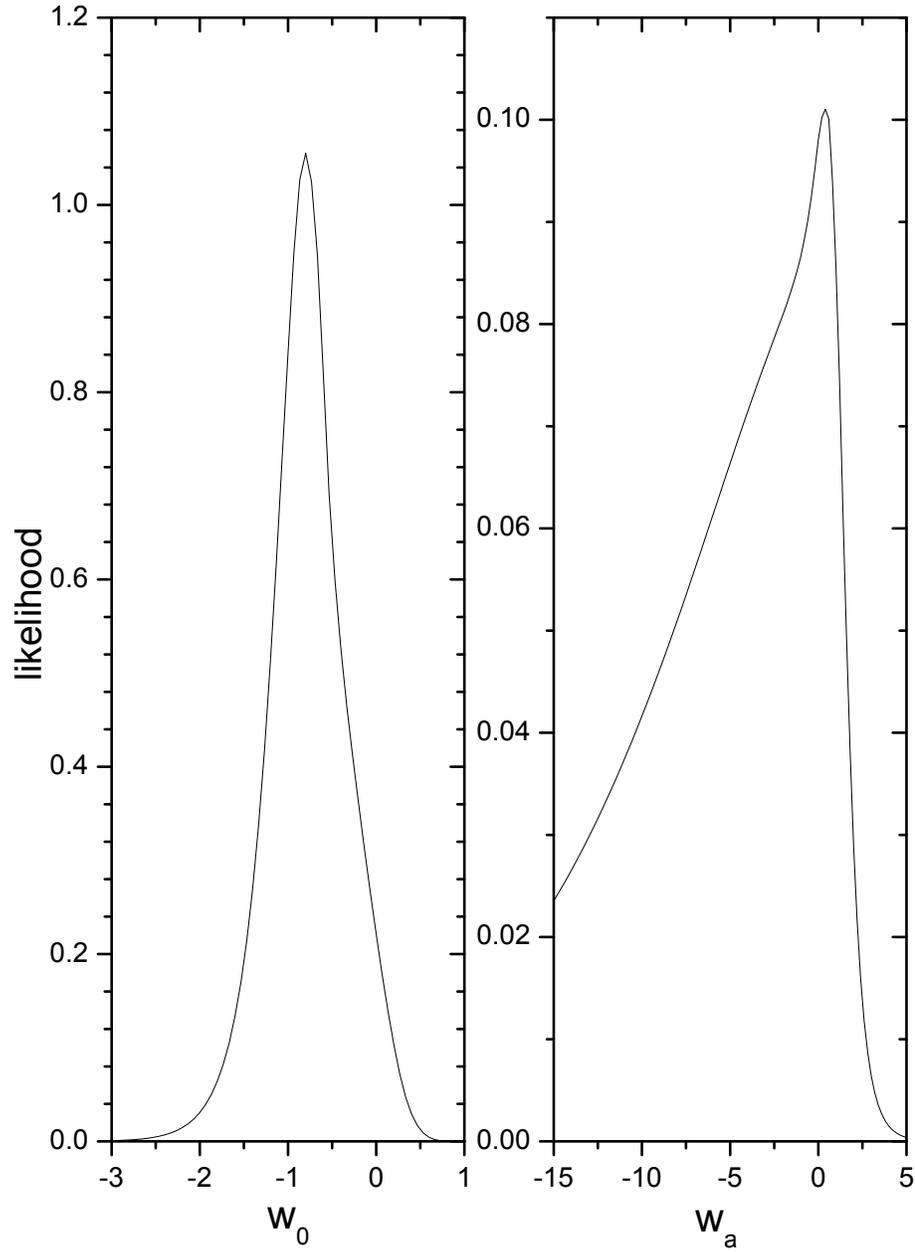}
\figcaption{\small The likelihoods of the parameters $w_0$ and
$w_a$. The maximum likelihood points are located at $w_0=-0.8$ and $
w_a = 0.4$, respectively. It can be seen that the likelihood of
parameter $w_a$ has a high non-Gaussian distribution.
\label{fig:SNe7} }
\end{figure}

We now focus on the time-varying model $w(z) = w_0 + w_a \, z /
(1+z)$. Figure \ref{fig:SNe4} shows the contours of ($w_0, w_a$),
the best-fit point is ($w_0, w_a)=(-0.73,0.84)$. It is seen that the
SNe Ia data alone have a poor constraint power on the parameter
$w_a$. In figure \ref{fig:SNe5}, we plot the contours of two
parameters ($\Omega_M, w_0$), the best-fit point is found to be
($\Omega_M, w_0)=(0.45,-0.68)$. It is shown that when $\Omega_M$
increases from $0.3$ to $0.45$, the allowed region of parameter
$w_0$ is enlarged quickly. For a smaller $\Omega_M < 0.3$, it leads
to  a much better constraint for the parameter $w_0$: $w_0 \sim -1.4
\sim -0.6$. Figure \ref{fig:SNe6} gives the contours of ($\Omega_M,
w_a$), the best-fit point is ($\Omega_M, w_a)=(0.44,-4.63)$. It is
noticed that when $\Omega_M$ increases from $0.34$ to $0.5$, the
allowed region for the  parameter $w_a$ is enlarged rapidly. For a
smaller $\Omega_M < 0.34$, it also leads to a much better constraint
for the parameter $w_a$: $w_a \sim - 3.0 \sim 2.5$. From the figure
\ref{fig:SNe5} and figure \ref{fig:SNe6}, it indicates that  an
extra restriction on $\Omega_M$ is necessary to improve the
constraint of the SNe Ia data on the parameters $w_0$ and $w_a$.
Figure \ref{fig:SNe7} shows the likelihoods of parameters $w_0$ and
$w_a$, in which the maximum likelihood points are located at
$w_0=-0.8$ and $ w_a = 0.4$, respectively. It can be seen that the
parameter $w_0$ is limited in the region $-2.5 <w_0 <0.5$ and the
likelihood of parameter $w_a$ has a high non-Gaussian distribution.

\section{JOINT ANALYSIS OF SNe Ia DATA, BAO AND SGL STATISTIC}

As we  have shown in the previous chapter, over $300$ SNe Ia
observed so far are not sufficient for determining the
cosmological parameters, especially for $w_0$ and $w_a$. Many
surveys(e.g. the Dark Energy Survey and Pan-STARRS) are proposed
to obtain the SNe Ia sample with enlarged number and improved
precision. Here we are going to constrain the dark energy through
the combination of the SNe Ia data, the baryon acoustic
oscillations as well as the  SGL splitting angel statistic.

\subsection{Baryon Acoustic Oscillations}

In the relativistic plasma of the early universe, ionized hydrogens
(protons and electrons) are coupled with energetic photons by
Thomson scattering. The plasma density is  uniform except for the
primordial cosmological perturbations. Driven by high pressure, the
plasma fluctuations spread outward at over half the speed of light.
After about $10^5$ years, the universe has cooled enough and the
protons capture the electrons to form neutral Hydrogen. This
decouples the photons from the baryons, which dramatically decreases
the sound speed  and effectively ends the sound wave propagation.
Because the universe has a significant fraction of baryons, these
baryon acoustic oscillations leave their imprint on very large scale
structures (about $100 {\rm Mpc}$) of the Universe.

The measurement of baryon acoustic oscillations was first processed
by the Sloan Digital Sky Survey (SDSS; York et al. 2000) and
Eisenstein et al. (2005) studied the large-scale correlation
function of its sample, which is composed of $46,748$ luminous red
galaxies over 3816 square degrees and in the redshift range $0.16$
to $0.47$. The typical redshift of the sample is at $z = 0.35$. The
large-scale correlation function is a combination of the
correlations measured in the radial (redshift space) and the
transverse (angular space) direction (Davis et al. 2007). Thus, the
relevant distance measure is modeled by the so-called dilation
scale, $D_V(z) = [D^2_A(z) z / H(z)]^{1/3}$, with comoving angular
diameter distance $D_A(z) = \int_0^z dz'/H(z')$. The dimensionless
combination $A(z) = D_V (z) \sqrt{\Omega_M H_0^2}/z$ has no
dependence on the Hubble constant $h$ and is found to be well
constrained by the SDSS data at $z=0.35$. A standard ruler is
provided as (Eisenstein et al. 2005)
 \beq
A = \frac{\sqrt{\Omega_M}}{[H(z_1)/H_0]^{1/3}}~\bigg\lbrack
~\frac{1}{z_1}~\int_0^{z_1}\frac{ dz}{H(z)/H_0}
~\bigg\rbrack^{2/3} = ~0.469  \pm 0.017~, \eeq where $z_1 = 0.35$.
This ruler is a $\Omega_M$ prior and can be used to constrain dark
energy. Then the statistic is given by
 \beq
 \chi^2(\Omega_M, w) = {[A(\Omega_M, w) - 0.469]^2 \over 0.017^2 }
 \label{eq:baochi2}
 \eeq

\subsection{SGL Splitting Angle Statistic}

The CLASS statistical sample has provided a well-defined statistical
sample with  $N=8958$ sources. Totally  $N_l=13$
 multiple image gravitational lenses have been discovered and all have image
separations $\Delta\theta<3^{\prime\prime}$ (Browne et al. 2003).
The SGL statistics are sensitive to the equation-of-state
parameter $w$ of dark energy, which influences the number density
of lens galaxies and the distances between the sources and lens.
The probability with image separations larger than $\Delta\theta$
for a source at redshift $z_s$ on account of the galaxies
distribution from the source to the observer can be obtained by
(Schneider et al. 1992)
 \beq
 P(>\Delta\theta) = \int_0^{z_s}\int_0^{\infty} {d D_L\over dz}
 (1+z)^3 n(M,z)\sigma(>\Delta\theta)\ dM dz\,,
 \label{eq:lp} \eeq
where $M$ is the mass of a dark halo, $D_L$ is the proper distance
from the observer to the lens, n(M, z) is the comoving number
density of dark halos virialized by redshift $z$ with mass $M$ and
$\sigma$ is the cross section for two images with a splitting angle
$ > \Delta\theta$.

According to the Press-Schechter theory, the comoving number density
with mass in the range $(M,M+dM)$ is given by
\begin{eqnarray}
     n(M,z)\,dM = {\rho_0\over M}\, f(M,z)\, dM\,. \label{eq:p-s0}
\end{eqnarray}
with the  matter density of universe today $\rho_0 = \Omega_M
\rho_{crit,0}$ and the critical matter density at present
$\rho_{{\rm crit},0}=3 H_0^2/(8\pi G)$.
 $f(M,z)$ is the Press-Schechter function, and we shall utilize the modified
 form by Sheth and Tormen (1999)
 \begin{eqnarray}
     & & f(M,z) = - {0.383\over \sqrt{\pi}}  {\delta_c\over \Delta^2} {d\Delta\over dM}
 \left[1+\left({\Delta^2\over0.707 \delta_c^2 }\right)^{0.3} \right] \times {\rm
 exp} \left[- {0.707\over2} \left({\delta_c\over
 \Delta}\right)^2\right]\,,
     \label{eq:p-s} \\
 & & \Delta^2 (M,z) = \int_0^{\infty} {dk\over k} \Delta_k(k,z) W^2(kr)
\end{eqnarray}
where $\Delta$ is the variance of the mass fluctuations(Eisenstein
and Hu 1999) and parameter $\delta_c (z)$ is the linear overdensity
threshold for a spherical collapse(Wang and Steinhardt 1998;
Weinberg and Kamionkowski 2002).

For different density profile of dark halo, the lensing cross
section $\sigma$ can be calculated out based on the lensing
equation. We shall use the combined mechanism of SIS and NFW model
to explain the whole experimental curve of strong gravitational
lensing statistic. For that a new model parameter $M_c$ was
introduced by Li and Ostriker (2002): lenses with mass $M<M_c$ have
the SIS profile, while lenses with mass $M > M_c$ have the NFW
profile. Then the differential probability is given by
 \[ dP/dM =
dP_{SIS}/dM\, \vartheta(M_c - M) + dP_{NFW}/dM\,\vartheta(M -
M_c)\] where $\vartheta$ is the step function, $\vartheta(x-y)=1$,
if $x>y$ and 0 otherwise. As the splitting angle $\Delta \theta$
is directly proportional to the mass $M$ of lens halos, the
contribution to large $\Delta \theta$ of SIS profile is depressed
by $M_c$. The lens data require a mass threshold $M_c \sim
10^{13}h^{-1}M_{\odot}$, which is consistent with the halo mass
whose cooling time equals the age of the universe today.

The likelihood function of the SGL splitting angle statistic is
defined as \beq {\rm L}(w) = (1-p(w))^{N-N_l}\prod_{i=1}^{N_l}
q_i(w). \label{eq:likh}\eeq $p(w)$ and $q_i(w)$ represent the
model-predicted lensing probabilities and the differential lensing
probabilities, respectively. They are related to $P$ in Equation
(\ref{eq:lp}) by an integration \begin{eqnarray}
   p(w) \equiv P_{\rm obs}(>\Delta\theta ) = \int\int B\, {d P(>\Delta\theta )\over
       dz}\, \varphi(z_s) dz dz_s\,,
    \label{eq:pobs}
\end{eqnarray}
and
\begin{eqnarray}
   q(w) \equiv {dP_{\rm obs}(>\Delta\theta ) \over d\Delta \theta } = \int\int B\, {d^2 P(>\Delta\theta )\over
       d\Delta\theta  dz}\, \varphi(z_s)  dz dz_s\,.
    \label{eq:pobs2}
\end{eqnarray}
${\rm B}$ is the magnification bias and can be found in our previous
work(Zhang et al. 2009). $\varphi(z_s)$ is the redshift distribution
of sources. Here we  take the Gaussian model by directly fitting the
redshift distribution of the subsample of CLASS statistical sample
provided by Marlow et al. (2000), which is given by (Zhang et al.
2009)
 \beq
 g(z_s) = {N_s \over \sqrt{2 \pi} \lambda } {\rm exp} \left[- {(z_s-a)^2}\over 2
 \lambda^2\right],
 \eeq with $N_s=1.6125; \; a=0.4224; \; \lambda=1.3761$.

\subsection{Joint Analysis and Numerical Results}

In this section, we will investigate the constraint on the
cosmological parameters from the joint analysis of (SNe + BAO) and
(SNe + BAO + SGL), respectively. For the two(or three) independent
observations, the likelihood function of a joint analysis is just
given by
 \beq
  L &=& L_{\rm SNe} \times L_{\rm BAO} \;\; (\times L_{\rm SGL})   \nonumber \\
    &=& \exp(-\chi_{\rm SNe}^2/2) \times \exp(-\chi_{\rm BAO}^2/2) \;\;  (\times L_{\rm SGL}).
 \eeq The statistic significance $\chi^2_{\rm BAO}$ and $\chi_{\rm SNe}^2$
can be obtained by using Equations (\ref{eq:baochi2}) and
(\ref{eq:snechi2}), respectively.  $L_{\rm SGL}$ is the likelihood
function of SGL statistic and can be obtained by using Equation
(\ref{eq:likh}).  The parameter $h$ in the SNe Ia data and SGL data
has different meaning and For the joint analysis of (SNe + SGL), the
parameter $h$ should be integrated separately: for the former, we
integrate $h$ in the range around $0.65$ and for the latter,  the
integral range is $0.4$ to $0.9$. The BAO analysis has no dependence
on the Hubble constant $h$.

\begin{figure}
\epsscale{0.8} \plotone{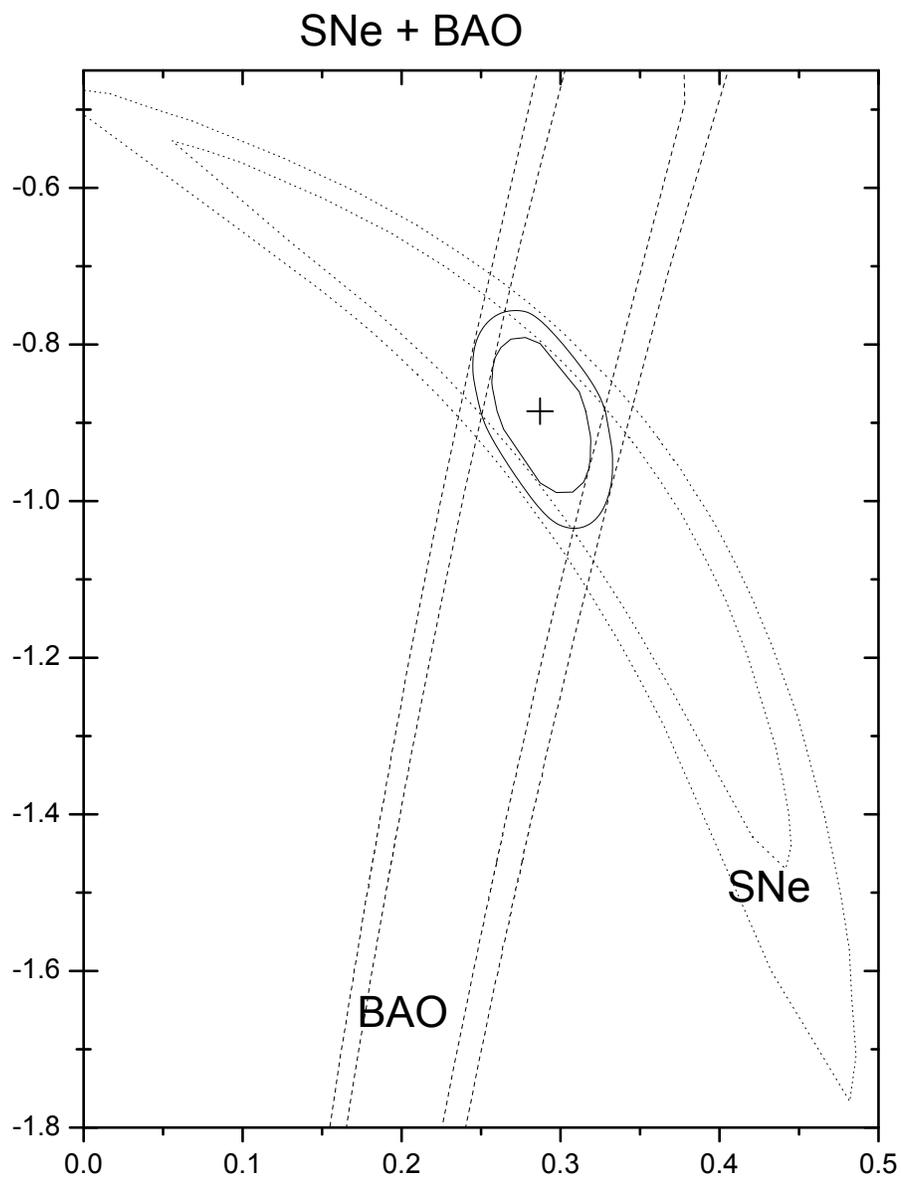}
\figcaption{\small  68\% C.L. and 95\% C.L. allowed regions of
($\Omega_M, w$) from the joint analysis of (SNe + BAO). The best-fit
result is $(\Omega_M, w) = (0.28, -0.88)$. \label{fig:SNe8} }
\end{figure}

Let us first discuss the constraints for the constant $w$ case from
the joint analysis of (SNe + BAO). After the Powell
minimization(Press, et al 1992), we get the best fit results of the
three parameters $(\Omega_M, w, h) = (0.287, -0.885, 0.646)$. The
fitted $\Omega_M$ decreases in comparison with the result of SNe
data alone, and at the same time the fitted $w$ increases due to
their degeneracy relation. Then in figure \ref{fig:SNe8}, we show
the constraints on $\Omega_M$ and the constant $w$ from the joint
analysis of (SNe + BAO). The parameter $h$ has been marginalized.
The best-fit results are $(\Omega_M, w) = (0.28, -0.88)$. The $95\%$
C.L. allowed regions of constant $w$ and $\Omega_M$ are found to be:
$-1.03\leq w \leq-0.75$ and $0.24 \leq \Omega_M \leq 0.33$,
respectively. It is seen that: (a) the most allowed region of
parameter $w$ is above the line $w=-1$; (b) the fitting results are
consistent with the $\Lambda {\rm CDM}$ cosmology at $95\%$ C.L..
Comparing with the results of SNe Ia data alone, we see that the
allowed regions for $\Omega_M$ and $w$ are much reduced and the
degeneracy between them disappears.

\begin{figure}
\epsscale{0.8} \plotone{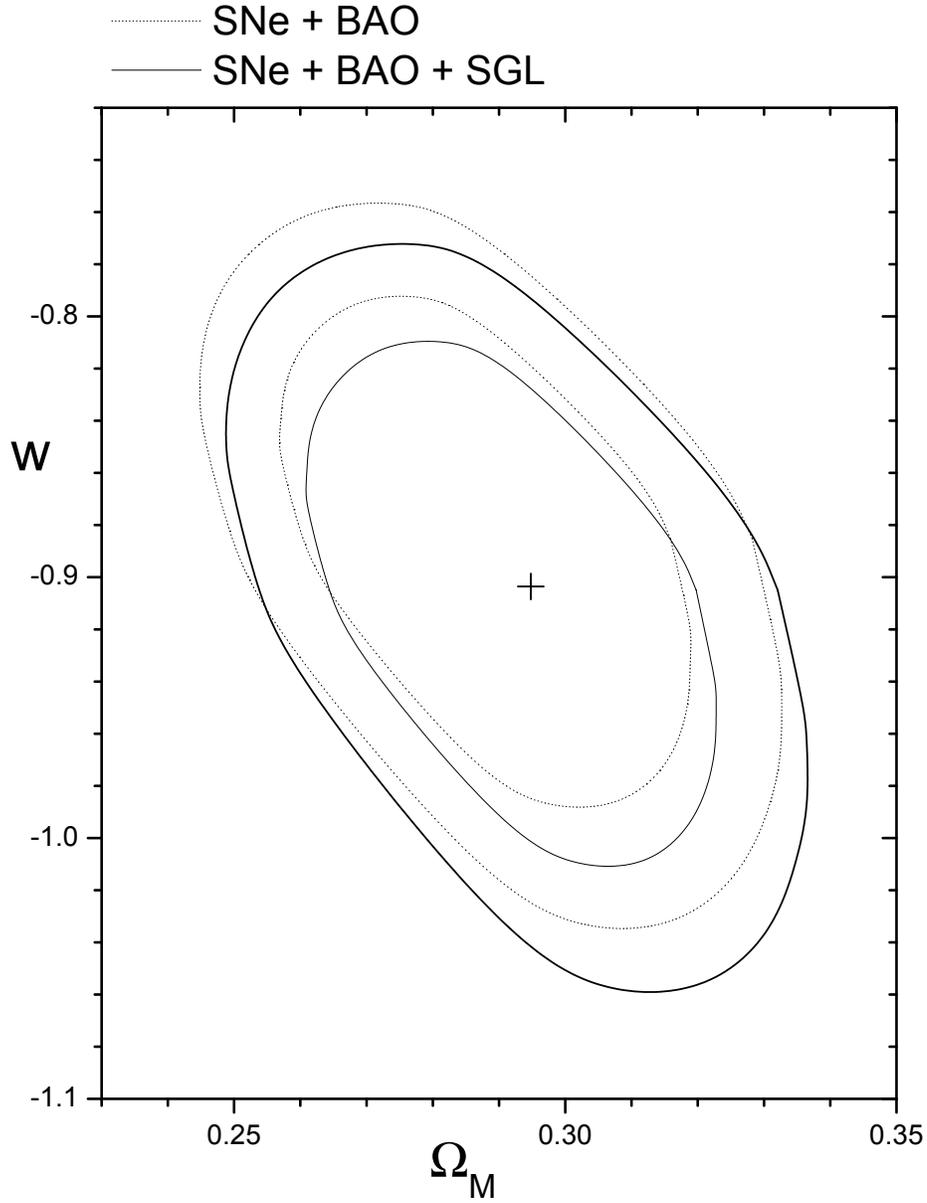}
\figcaption{\small  68\% C.L. and 95\% C.L. allowed regions of
($\Omega_M, w$) from the joint analysis of (SNe + BAO + SGL) (solid
lines) in comparison with the joint analysis of (SNe + BAO)(dotted
lines). The best-fit result from (SNe+ BAO + SGL) is $(\Omega_M, w)
= (0.29, -0.91)$. \label{fig:SNe9} }
\end{figure}

In figure \ref{fig:SNe9}, we show the constraints on $\Omega_M$ and
the constant $w$ from the joint analysis of (SNe + BAO + SGL). The
parameter $h$ has been marginalized again. The best fit result is
$(\Omega_M, w) = (0.29, -0.91)$. The $95\%$ C.L. allowed regions of
constant $w$ and $\Omega_M$ are found to be: $-1.06 \leq w \leq
-0.77$ and $0.25 \leq \Omega_M \leq 0.34$. Comparing with the
results of (SNe + BAO) case, it is seen that the results have only
slight differences and the fitted $w$ is found to be slightly
smaller after adding the SGL data.

\begin{figure}
\epsscale{0.8} \plotone{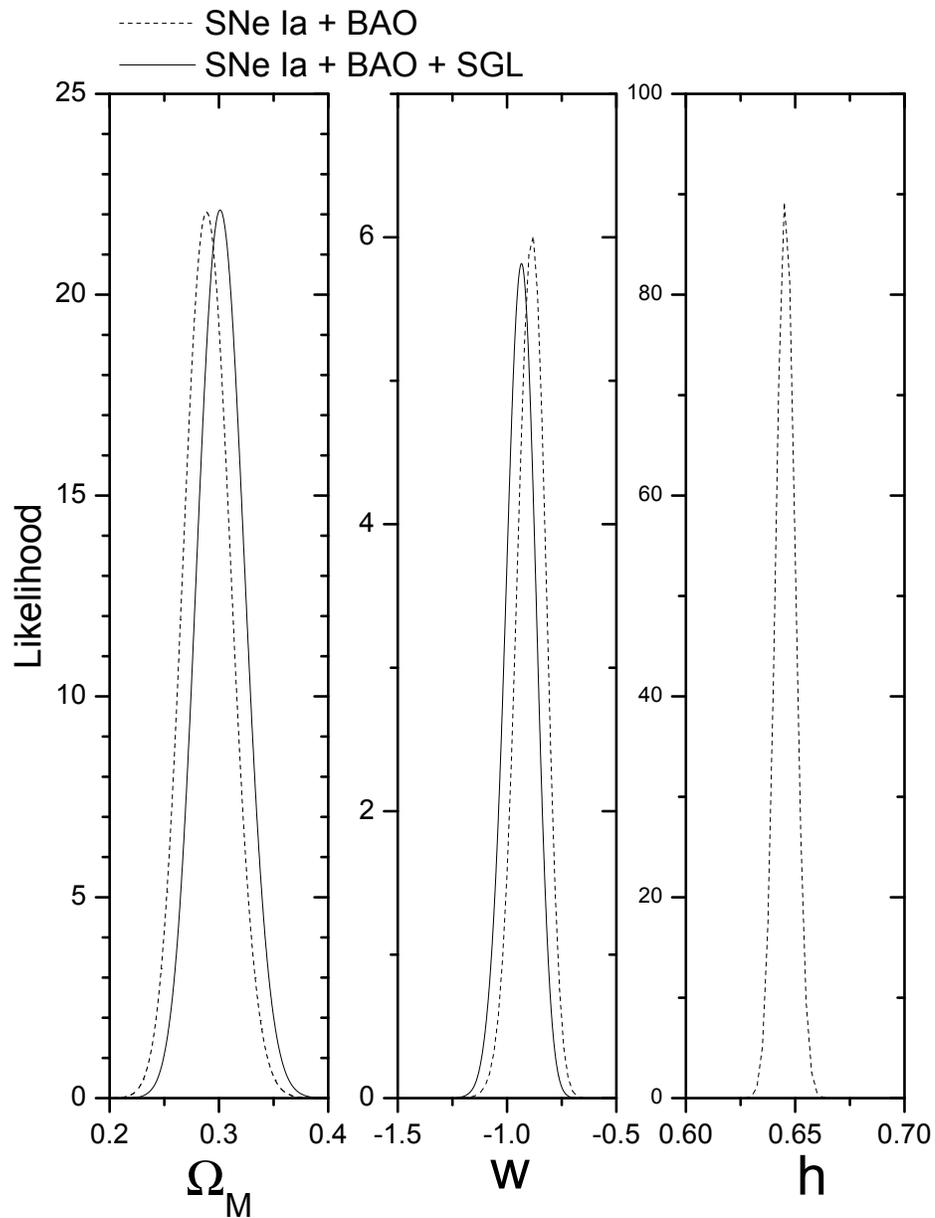}
\figcaption{\small   The likelihoods of the parameters $\Omega_M$,
$w$ and $h$ from the joint analysis of (SNe + BAO) and (SNe + BAO +
SGL), respectively. The maximum likelihood points are located at
$\Omega_M=0.29$, $w=-0.88$ and $ h = 0.65$ for (SNe + BAO) and
$\Omega_M=0.296$ and $w=-0.91$ for (SNe + BAO + SGL).
\label{fig:SNe10} }
\end{figure}

Figure \ref{fig:SNe10} plots the likelihoods for the parameters
$\Omega_M$, $w$ and $h$ from the joint analysis of (SNe + BAO) and
for the parameters $\Omega_M$ and $w$ from (SNe + BAO + SGL),
respectively. The maximum likelihood points are located at
$\Omega_M=0.29$, $w=-0.88$ and $ h = 0.65$ for (SNe + BAO) and
$\Omega_M=0.296$ and $w=-0.91$ for (SNe + BAO + SGL). It is
interesting to find that the parameters $w$ and $\Omega_M$ are
restricted to the range: $-1.17\leq w \leq -0.67$ and $0.23 \leq
\Omega_M \leq 0.37$.

\begin{figure}
\epsscale{0.8} \plotone{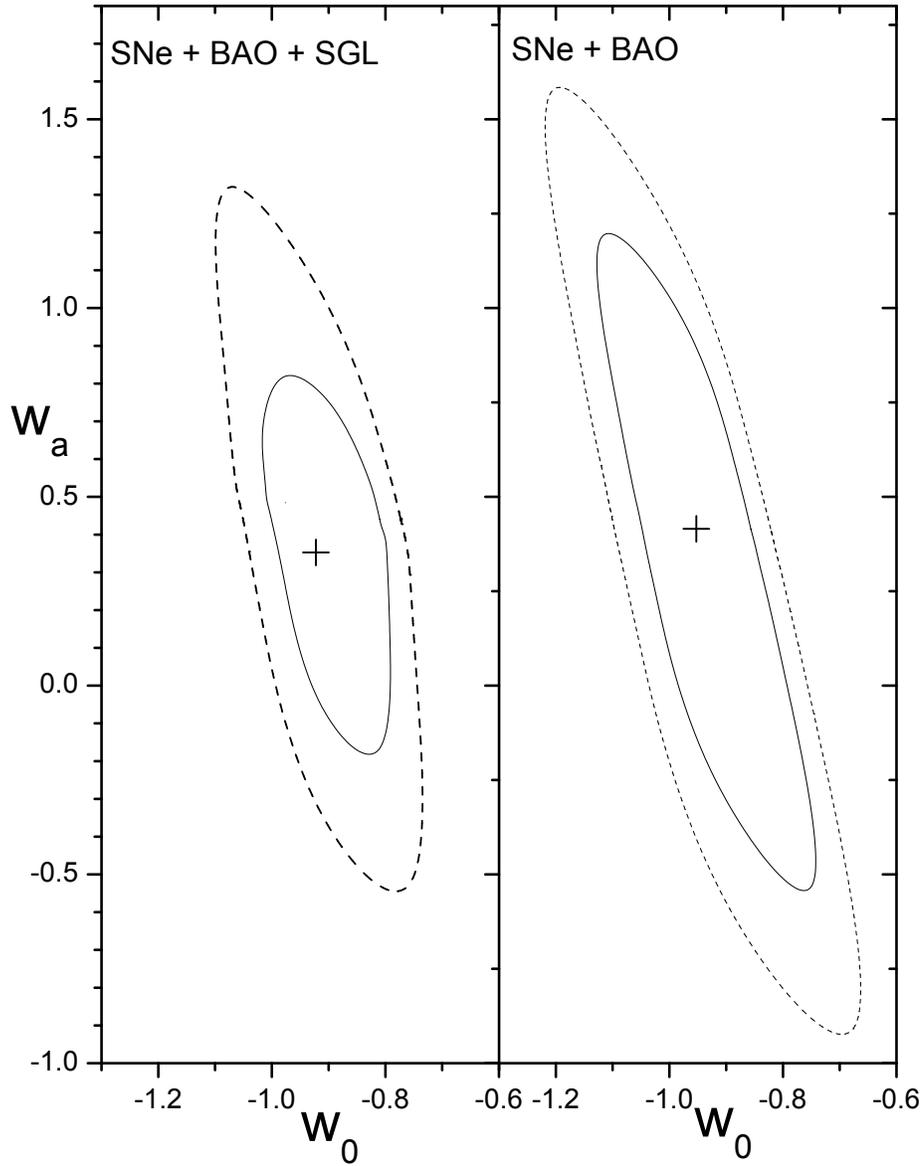}
\figcaption{\small the 68\% C.L. and 95\% C.L. allowed regions of
($w_0, w_a$) from the joint analysis of (SNe + BAO) and (SNe + BAO +
SGL), respectively. The crosshairs mark the best-fit point $(w_0,
w_a) = (-0.95, 0.41)$ for the (SNe + BAO) case and $(w_0,
w_a)=(-0.92, 0.35)$ for the (SNe + BAO + SGL) case.
\label{fig:SNe13}}
\end{figure}

After marginalizing the cosmological parameters $(\Omega_M, h)$, we
obtain the constraint on $(w_0, w_a)$ in figure \ref{fig:SNe13} from
the joint analysis of (SNe + BAO) and (SNe + BAO + SGL),
respectively. The crosshairs mark the best-fit point $(w_0, w_a) =
(-0.95, 0.41)$ for the (SNe + BAO) case and $(w_0, w_a)=(-0.92,
0.35)$ for the (SNe + BAO + SGL) case. For the (SNe + BAO) case, the
$95\%$ C.L. allowed regions for the parameters $w_0$ and $w_a$ are
found to be: $-1.22 \leq w_0 \leq -0.66$ and $-0.92 \leq w_a \leq
1.59$. For the (SNe + BAO + SGL) case, the $95\%$ C.L. allowed
regions for the parameters $w_0$ and $w_a$ are found to be:
$-1.10\leq w_0 \leq -0.72$ and $-0.55\leq w_a \leq 1.32$. After
adding the SGL data, the constraint on the parameter $w_0$ is
improved moderately, but for the parameter $w_a$, the allowed region
decreases by near half. The extra constraint power on the
time-varying $w(z)$ obtained through adding SGL data is due to the
larger redshift $0 < z < 3.0$ of the galaxies in CLASS observational
sample, in comparison with the reshift range of SNe data $0 < z <
1.5$ and the redshift of BAO $z=0.35$. It can be seen for the both
cases that: (a) the most allowed region of $w_a$ is above $w_a=0$;
(b) in comparison with the cosmological constant ($w_0, w_a) =(-1.0,
0.0)$, the joint analysis for both cases favors more positive ($w_0,
w_a$); (c) in comparison with the results of SNe Ia data alone, the
constraint on $w_a$ is much improved and $w_0$ also gets better
constrained.

\begin{figure}
\epsscale{0.8} \plotone{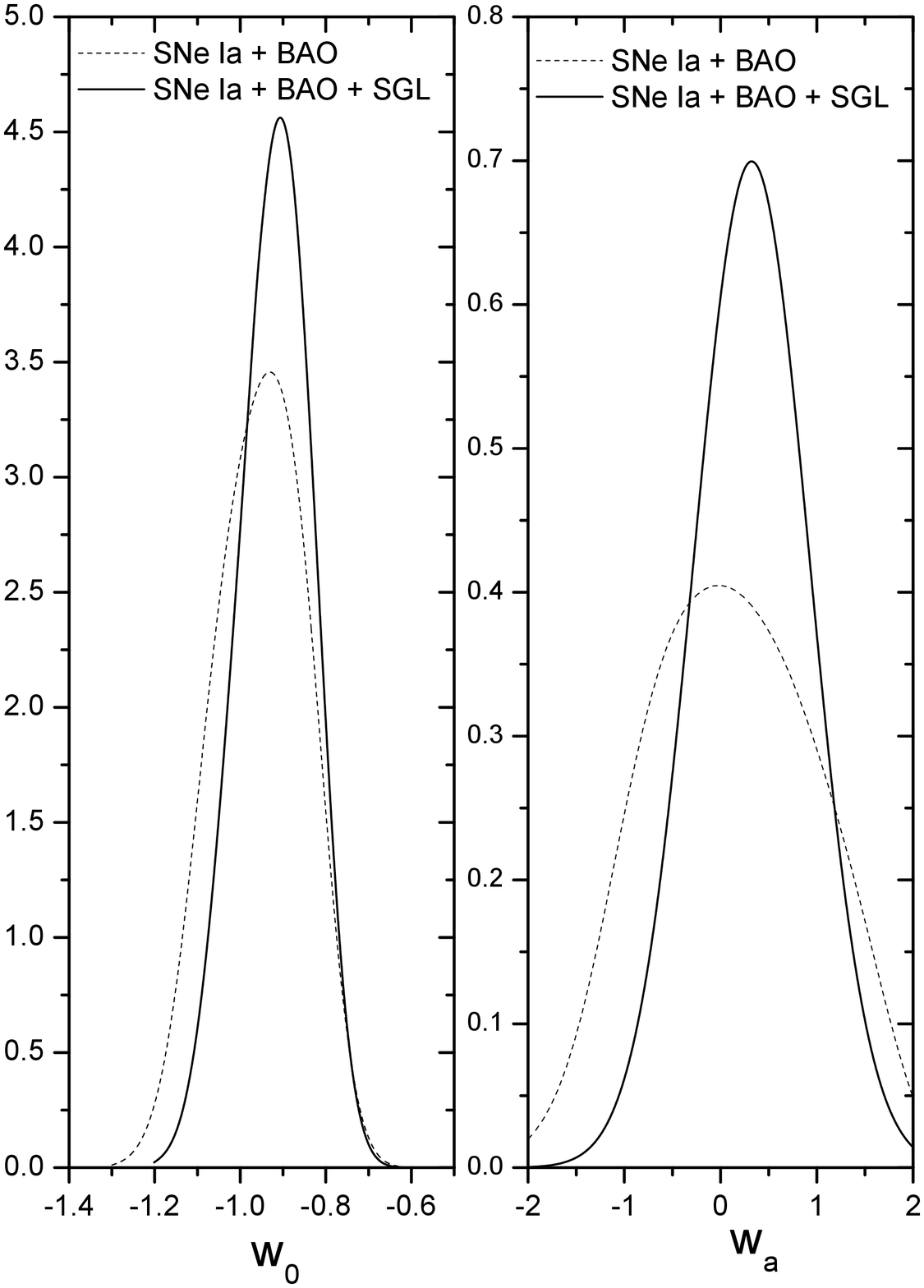}
\figcaption{\small the likelihoods of parameters $w_0$ and $w_a$
from the joint analysis of (SNe + BAO) and (SNe + BAO + SGL),
respectively. For (SNe + BAO) case, the maximum likelihood points
are located at $w_0=-0.94$ and $w_a = 0.0$, respectively. For (SNe +
BAO + SGL) case, the maximum likelihood points are $w_0=-0.91$ and
$w_a = 0.34$, repectively.\label{fig:SNe14}}
\end{figure}

Figure \ref{fig:SNe14} plots the likelihoods of parameters $w_0$ and
$w_a$ from the joint analysis of (SNe + BAO) and (SNe + BAO + SGL),
respectively. For (SNe + BAO) case, the maximum likelihood points
are located at $w_0=-0.94$ and $w_a = 0$. Note that $w_a=0$ implies
a constant equation-of-state of dark energy. For (SNe + BAO + SGL)
case, the maximum likelihood points are found to be $w_0=-0.91$ and
$w_a = 0.34$. We see that the parameters $w_0$ and $w_a$ are
restricted to be: $-1.20\leq w_0 \leq -0.67$ and $-1.0 \leq w_a \leq
2.0$.

\section{CONCLUSIONS}

We have carefully investigated, based on the latest SNe Ia data, BAO
and SGL statistic, the constraint on the equation-of-state parameter
$w$ of dark energy for both constant and time varying cases in the
flat cosmology.  The influences of Hubble constant h and matter
density $\Omega_M$ on the fitting results  are carefully
demonstrated. The typical redshift measured by the three kinds of
observations is $z \sim 1 $ and far smaller than the redshift of CMB
involved, their constraints on the parameter $w$ are effective and
significant only for the redshift region $z < 1.5$.

The influence of the equation-of-state parameter $w$ on the density
$\rho(z)$ of dark energy in the universe and the distance $d(z)$
makes SNe Ia data a powerful probe of dark energy. In this paper, we
have utilized the latest $324$ SNe Ia data provided by Hicken et al.
(2009) using MLCS17 light curve fitter with the best cuts $A_V \leq
0.5$ and $\Delta < 0.7$ to carefully investigate the constraint on
the constant $w$ and especially the time-varying
$w(z)=w_0+w_az/(1+z)$. The influences of Hubble constant $h$ and
matter density $\Omega_M$ have also been studied carefully. For the
constant $w$ case, we have shown that: (a) the best-fit results for
the three correlated parameters are found to be $(\Omega_M, w,
h)=(0.358, -1.09, 0.647)$, it is seen that $\Omega_M$ is somewhat
large in comparison with $\Omega_M = 0.26$ of the concordance
cosmology provided by WMAP five year data(Komatsu et al. 2009); note
that using a different parameterization of dark energy, Huang et al.
(2009) presented a best-fitted $\Omega_M=0.446$ from SNe Ia data,
which is even larger but still consistent with our result at 95\%
C.L.; (b) the fitted results of both $(\Omega_M,w)$ from the SNe Ia
data is very sensitive to the value of Hubble constant $h$: when $h$
increases, the fitted $\Omega_M$ increases and $w$ decreases very
rapidly; (c) for parameter $\Omega_M$, the SNe Ia data alone can
only give a large allowed region $0.0 \leq \Omega_M \leq  0.5$ at
95\% C.L.; (d) for the constant $w$ case, the likelihoods are found
to be: $w = -0.88^{+0.31}_{-0.39}$ and
$\Omega_M=0.36^{+0.09}_{-0.15}$ after marginalizing other parameters
in obtaining each of them, which is consistent with the $\Lambda \rm
CDM$ at $95\%$ C.L.; (e) there is clear degeneracy of constant $w$
and $\Omega_M$, which depresses the power of SNe Ia to constrain two
parameters. It has been shown that the parameter $w$ is limited from
$-2.0$ to $-0.5$ and $\Omega_M$ should be less than $0.5$.

In particular, we have paid special attention to the constraints on
the time-varying case parameterized by two parameters ($w_0, w_a$).
After marginalizing the parameters $\Omega_M$ and $h$, we have
obtained the fitting results $(w_0, w_a)=(-0.73^{+0.23}_{-0.97},
0.84^{+1.66}_{-10.34})$, which indicates that (a) the SNe Ia data
alone have only a poor constraint power on the parameter $w_a$, an
extra restriction of $\Omega_M$ is necessary, so that the constraint
of SNe Ia on the parameters $w_0$ and $w_a$ can be much improved;
(b)the likelihood of parameter $w_a$ has a high non-Gaussian
distribution.

The summary parameter of BAO can provide a standard ruler by which
the absolute distance of $z=0.35$ can be determined with $5\%$
accuracy. This ruler can be a $\Omega_M$ prior and has been used to
constrain dark energy. The strong gravitational lensing (SGL)
statistic is a useful probe of dark energy.  Through comparing the
observed number of lenses with the theoretical expected result, it
enables us to constrain the parameter $w$. We have used the latest
SNe Ia data together with  the BAO (and the CLASS statistical
sample) to constraint dark energy. For the constant $w$ case, the
results obtained from (SNe + BAO) and (SNe + BAO + SGL) only have a
slight difference: (a) for the (SNe + BAO) case, the best fit
results of the three parameters $(\Omega_M, w, h)$ are $(0.287,
-0.885, 0.65)$  and for the (SNe + BAO + SGL) case, the best fit
point is $(\Omega_M, w) = (0.298, -0.907)$ where $h$ has been
marginalized; (b) the fitting results of the parameter $\Omega_M$
are found to be $\Omega_M=0.29^{+0.02}_{-0.03}$ for the (SNe + BAO)
case and $\Omega_M = 0.29^{+0.03}_{-0.03}$ for the (SNe + BAO + SGL)
case; (c) the fitting results for the constant $w$ case are found to
be $w = -0.88^{+0.07}_{-0.09}$ for the (SNe + BAO) case and $w =
-0.91^{+0.10}_{-0.10}$ for the (SNe + BAO + SGL) case, which are
consistent with the $\Lambda \rm CDM$ at 95\% C.L.; (d) the most
allowed region of parameter $w$ is above the line $w=-1$. Comparing
with the fitting results from the SNe Ia data alone, we have found:
(a) the allowed region at 95\% C.L. for $\Omega_M$ is reduced to
one-fifth; (b)  the best fit value of $w$ is almost not changed but
its variance is reduced very much.

For the time-varying case $w(z)$  after marginalizing $(\Omega_M,
h)$, we have obtained the fitting results $(w_0,
w_a)=(-0.95^{+0.45}_{-0.18}, 0.41^{+0.79}_{-0.96})$ for the (SNe +
BAO) case and $(w_0, w_a)=(-0.92^{+0.14}_{-0.10},
0.35^{+0.47}_{-0.54})$ for the (SNe + BAO + SGL) case. It has been
seen that the adding of the SGL data makes the constraints on
parameter ($w_0, w_a$) to be much improved. For both cases, the most
allowed region of $w_a$ is above $w_a=0$, which indicates that the
phantom type models are disfavored. Comparing with the fitting
results from the latest SNe Ia data alone, we have observed that:
(a) the best fit values for $w_0$ are decreased by over $0.2$ and
the variances are approximately reduced to one-fourth; (b) the best
fit values of $w_a$ are decreased by $0.49$ and the variances are
reduced to one-twelfth.

In conclusion, the latest MLCS17 data set given by Hicken et al.
(2009) has provided an interesting constraint on the cosmological
parameters. The joint analysis of SNe Ia and BAO breaks the
degeneracy between $w$ and $\Omega_M$ and leads to a more stringent
to constrain on the dark energy and matter density than the SNe Ia
data alone, especially for the time-varying case with parameters
($w_0, w_a$). The adding of the SGL data can further improve the
constraint for ($w_0, w_a$). A large number of SNe Ia samples with
reduced systematical uncertainties in the near future, together with
possible new observations on BAO and SGL statistic, would be very
useful to understand the properties of dark energy and Hubble
constant.

\section*{Acknowledgments}
The author (YLW) would like to thank Q.G. Huang and M. Li for useful
discussions. This work was supported in part by the National Science
Foundation of China (NSFC) under the grant \# 10821504, 10491306 and
the Project of Knowledge Innovation Program (PKIP) of Chinese
Academy of Science.


\clearpage

\newpage
\begin{thebibliography}{}
\bibitem{Albrecht2006}
Albrecht, A. et al., 2006, arXiv:astro-ph/0609591
\bibitem{barger2005}
Barger, V. Guarnaccia, E and Marfatia, D. 2006, Phys.Lett. B635
61-65
\bibitem{biswas2009}
Biswas, R., and Wandelt, B. D., 2009, arXiv:0903.2532
\bibitem{branch1998}
Branch, D, 1998, Ann. Rev. Astron. Astrophys. 36, 17;
\bibitem{browne2003}
 Browne, I.A., et al. 2003, MNRAS, 341, 13
\bibitem{Chae2007}
Chae, K.-H. 2007, ApJ, 658, 71
\bibitem{david} Chevallier, M., and Polarski, D.
Int. J. Mod. Phys. D10, 213 (2001)
\bibitem{quiteen}
Caldwell, R. R., Dave, R., and Steinhardt, P. J. 1998, Phys. Rev.
Lett., 80, 1582
\bibitem{caldwell2002}
Caldwell, R. R., Phys. Lett. B 545, 23(2002)
\bibitem{davis}
Davis, T. M. et al. 2007, ApJ, 666, 716
\bibitem{eisen99} Eisenstein, D. J., and Hu, W. 1999, ApJ, 511, 5
\bibitem{eisenstein2005}
Eisenstein, D. J. et al., 2005, ApJ, 633, 560
\bibitem{gibson2001}
Gibson, B.K., and Brook, C.B. 2001, arXiv:astro-ph/0011567
\bibitem{guy2005}
Guy, J., Astier, P., Nobili, S., Regnault, N., and Pain, R. 2005,
AandA, 443, 781
\bibitem{guy2007}
Guy, J., et al. 2007, AandA 466, 11
\bibitem{Hicken2009}
Hicken, M. et al., 2009,  arXiv:0901.4804
\bibitem{Hinshaw2009}
Hinshaw, G.,  et al. 2009, ApJS, 180, 225
\bibitem{huang2009}
Huang, Q.G., Li, M., Li, X.D.,  and Wang, S. arXiv:0905.0797, 2009.
\bibitem{Huterer2004}
Huterer, D. and Ma, C. P., 2004, ApJ, 600, 7
\bibitem{jha2007}
Jha, S., Riess, A. G., and Kirshner, R. P. 2007, ApJ, 659, 122
\bibitem{Komatsu2009}
Komatsu, E., et al. 2009, ApJS, 180, 330
\bibitem{Kowalski2008}
Kowalski, M., et al. 2008, ApJ, 686, 749
\bibitem{li02}
Li, L. -X., and Ostriker, J.P. 2002, ApJ, 566, 652
\bibitem{linder2003}
Linder, E.V., Phys.Rev.Lett.90, 091301(2003)
\bibitem{marlow2000}
Marlow, D. R., Rusin, D., Jackson, N., Wilkinson, P. N., and Browne,
I. W. A., 2000, AJ, 119, 2629
\bibitem{cosconstan}
Padmanabhan, T. 2003, Phys. Rept. 380, 235
\bibitem{perlmutter}
Perlmutter, S. et al. 1999, ApJ, 517, 565
\bibitem{Porciani2000}
Porciani, C., and Madau, P. 2000, ApJ, 532, 679
\bibitem{press1992}
Press, W.H., Teukolsky, S.A., Vetterling, W.T.,  and Flannery, B.P.,
1992,  Numerical Recipes in Fortran (New York: Cambridge University
Press)
\bibitem{riess98}
Riess, A. G., et al. 1998b, AJ, 116, 1009
\bibitem{riess 04}
Riess, A. G., et al. 2004, ApJ, 607, 665
\bibitem{riess 07}
Riess, A. G., et al. 2007, ApJ, 659, 98
\bibitem{riess 09}
Riess, A. G., et al. 2009,  arXiv:0905.0697
\bibitem{riatra 88}
Ratra, B., and Peebles, P. J. E. 1988, Phys. Rev. D, 37, 3406
\bibitem{sandage1996}
Sandage, A., Saha, A., Tammann, G. A., Labhardt, L., Panagia, N.,
and Macchetto, F. D. 1996, ApJ, 460, L15
\bibitem{sandage2006}
Sandage, A.,  Tammann, G.A., Saha, A., Reindl,  B.,  Macchetto,
F.D., and Panagia, N., 2006, ApJ, 653, 843
\bibitem{sarbu2001}
Sarbu, N., Rusin, D., and Ma, C.-P. 2001, ApJ, 561, L147
\bibitem{schae1996}
Schaefer, B.E., 1996, ApJ, 459, 438
\bibitem{schne92}
Schneider, P., Ehlers, J., and Falco, E. E. 1992, Gravitational
Lenses (Berlin: Springer-Verlag)
\bibitem{sheth1999}
Sheth R. K., and Tormen G., 1999, MNRAS, 308, 119
\bibitem{wang1998}
Wang, L., and Steinhardt, P. J. 1998, ApJ, 508, 483
\bibitem{weinberg 2002}
Weinberg, N. N., and Kamionkowski, M. 2002, MNRAS, 337, 1269
\bibitem{wood2007}
Wood-Vasey, W. M. et al., 2007, ApJ, 666, 694w
\bibitem{york2007}
York, D.G., et al., 2000, AJ, 120, 1579
\bibitem{zhan2006}
Zhan, H. and  Knox, L.,  2006, arXiv:astro-ph/0611159
\bibitem{zhang2007}
Zhang, Q.J., Cheng, L.M. and Wu, Y.L.,  2009, ApJ, 694, 1402
\end{thebibliography}
\end{document}